\documentclass[aps,prl,twocolumn,amsmath,amssymb,showpacs,superscriptaddress]{revtex4-1}

\usepackage{graphicx}
\usepackage{color}

\makeatletter

\begin{document}

\title{Observation of Fermi Arcs in Type-II Weyl Semimetal Candidate WTe$_2$}

\author{Yun Wu}
\author{Na Hyun Jo}
\author{Daixiang Mou}
\author{Lunan Huang}
\author{S.~L.~Bud'ko}
\author{P. C. Canfield}
\email[]{canfield@ameslab.gov}
\author{Adam Kaminski}
\email[]{kaminski@ameslab.gov}
\affiliation{ Ames Laboratory, U.S. DOE and Department of Physics and Astronomy, Iowa State University, Ames, Iowa 50011, USA}

\date{\today}

\begin{abstract}
We use ultrahigh resolution, tunable, vacuum ultraviolet laser angle-resolved photoemission spectroscopy (ARPES) to study the electronic properties of WTe$_2$, a material that was predicted to be a type-II Weyl semimetal. The Weyl fermion states in WTe$_2$ were proposed to emerge at the crossing points of electron and hole pockets; and Fermi arcs connecting electron and hole pockets would be visible in the spectral function on (001) surface. Here we report the observation of such Fermi arcs in WTe$_2$ confirming the theoretical predictions. This provides strong evidence for type-II Weyl semimetallic states in WTe$_2$.
\end{abstract}


\maketitle

The discovery of graphene \cite{Geim07NatMat} opened possibility to study relativistic quasiparticles that can be realized in solids. Occurrence of  Dirac dispersion attracted great interests and trigered searches for novel topological states in three-dimensional (3D) systems \cite{Hasan10RMP, Qi11RMP}. Dirac semimetals with bulk 3D Dirac points protected by crystal symmetry have been proposed to exist in $\beta$-cristobalite BiO$_2$ \cite{Young12PRL} and A$_3$Bi (A = Na, K, Rb) \cite{Wang12PRB}. Na$_3$Bi and Cd$_3$As$_2$ were soon verified to be 3D Dirac semimetals with Dirac dispersion along all three directions in the momentum space \cite{Wang13PRB, Liu14Sci, Neupane14NatCom, Liu14NatMat, Yi14SciRep, Borisenko14PRL}. This lead to observation of novel topological quantum states with Fermi arcs \cite{Wan11PRB, Xu11PRL}, which were first observed in Na$_3$Bi \cite{Xu15SciObs}. Subsequently, another type of massless particle -- the Weyl fermion \cite{Weyl29ZfP}-- has been predicted to exist in a family of non-centrosymmetric transition metal monophosphides \cite{Huang15NatCom, Weng15PRX}. ARPES measurements on TaAs \cite{Xu15SciDis, Yang15NatPhys, Lv15NatPhys, Lv2015Experimental} and NbAs \cite{Xu15NatPhys} confirmed the existence of Fermi arcs connecting Weyl points of opposite chirality. Recently, a new type of Weyl semimetals (type-II Weyl semimetals) was proposed to possess Weyl points emerging at the boundary between electron and hole pockets \cite{Soluyanov2015Type}. WTe$_2$ \cite{Soluyanov2015Type} and MoTe$_2$ \cite{Sun15Prediction} were among the first predicted to be type-II Weyl semimetals with different Fermi arc length. By doping Mo in WTe$_2$, the Fermi arc length (or the topological strength) can be continuously tuned \cite{Chang15Arc}. Signatures of topological Fermi arcs have been reported in Mo doped WTe$_2$ by using pump laser to access the states above the Fermi level \cite{Belopolski2015Unoccupied}. Spectroscopic evidence for type-II Weyl semimetal states in MoTe$_2$ was reported and novel ``track states'' were predicted by theoretical modelling and density functional theory calculations \cite{McCormick2016Minimal} and subsequently discovered by ARPES \cite{Huang2016Spectroscopic}. In addition to  W(Mo)Te$_{2}$ family  \cite{Belopolski2015Unoccupied, Huang2016Spectroscopic, Deng2016Experimental, Jiang2016Observation, Liang2016Electronic, Xu2016Discovery}, YbMnBi$_{2}$ \cite{Borisenko2016Time} and LaAlGe \cite{XuSY2016Discovery} were also reported to display signatures of type-II Weyl semimetal states.

WTe$_2$ has attracted great interests due to its extremely large magnetoresistance at low temperatures and high magnetic fields \cite{Ali14Nat}. Other interesting phenomena have also been observed or proposed. Superconductivity has been reported to emerge from a suppressed magnetoresistive state by applying high pressure \cite{Kang15NatComm, Pan15NatComm}. Pressure induced quantum phase transition with the changes of Fermi surface structure, i.e., Lifshitz transition, was proposed \cite{Kang15NatComm}. Interestingly, temperature induced Lifshitz transition has also been observed in WTe$_2$ and the dramatic shifts of the chemical potential with temperature was attributed to the close proximity of electron and hole densities of states near the Fermi energy \cite{Wu2015PRL}. Type-II Weyl semimetal states have also been proposed to exist in WTe$_{2}$ \cite{Soluyanov2015Type}. However, the Weyl points reside at roughly 50 meV above the Fermi level and separated by just a few meV, making it difficult to be observed by ARPES measurements. Fortunately, the Weyl points project to distinct locations on the (001) surface and Fermi arcs should emerge connecting electron and hole pockets with opposite Chern number \cite{Soluyanov2015Type}. This distinct topological surface states can be easily observed by ARPES measurements. Although the band structures and Fermi surface of WTe$_2$ have been reported previously \cite{Pletikosic14PRL, Wu2015PRL, Jiang2015Signature}, no surface states were clearly observed. 

\begin{figure}[tb]
	\includegraphics[width=3.4in]{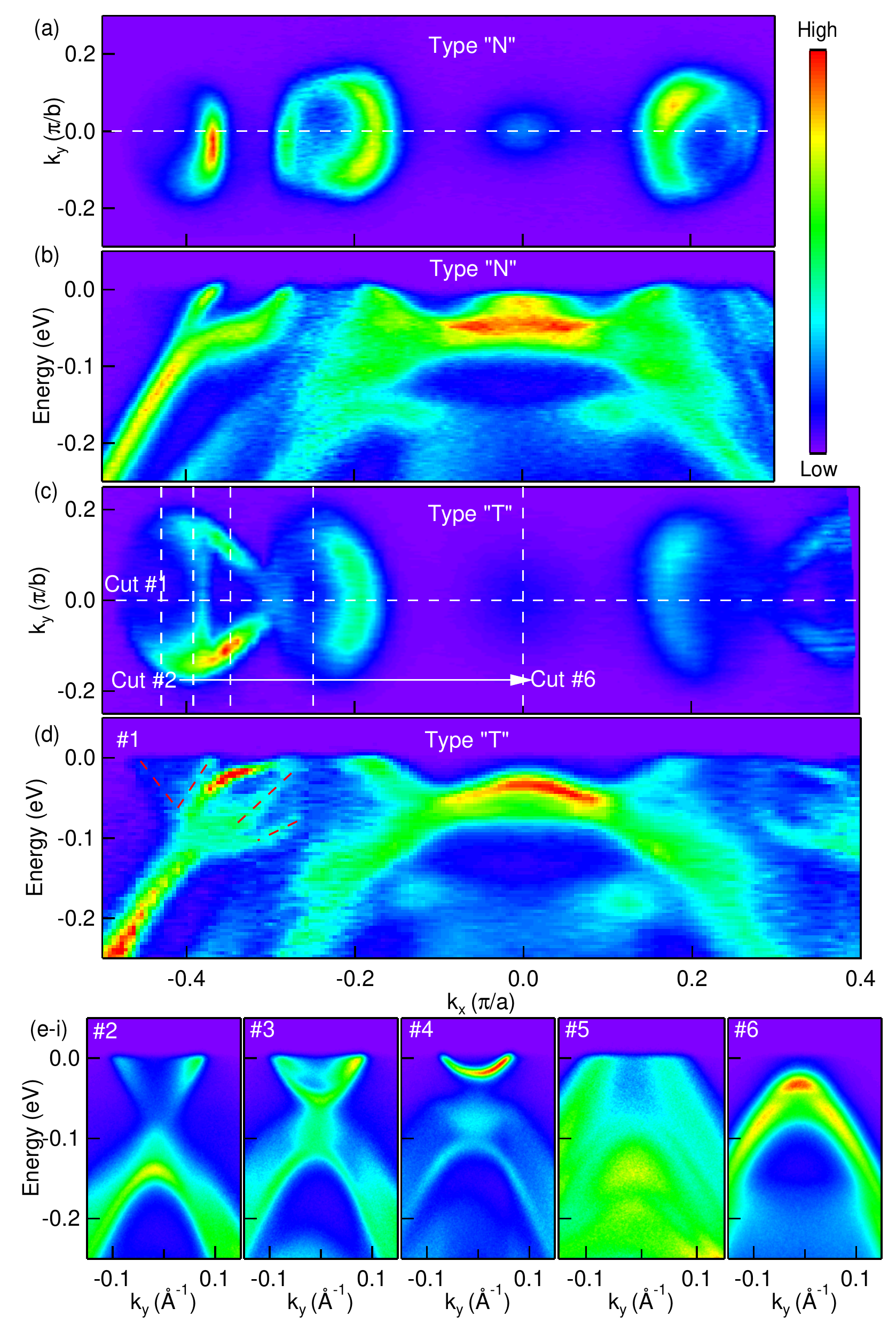}%
	\caption{The Fermi surface and band dispersion measured at photon energy of 6.7 eV for the two types of cleaves. 
	(a) ARPES intensity integrated within 10 meV about the chemical potential measured at $T=40$~K for type ``N" cleave.
	(b) Band Dispersion along white dashed line cut in (a). 
	(c) ARPES intensity integrated within 10 meV about the chemical potential measured at $T=16$~K for type ``N" cleave.
	(d)-(i) Band Dispersions along cuts \#1--\#6 in (c). Red dashed lines in (b) marked the electron pocket and two left branches of the hole bands. 
	\label{fig:Fig1}}
\end{figure}

\begin{figure}[tb]
	\includegraphics[width=3.4in]{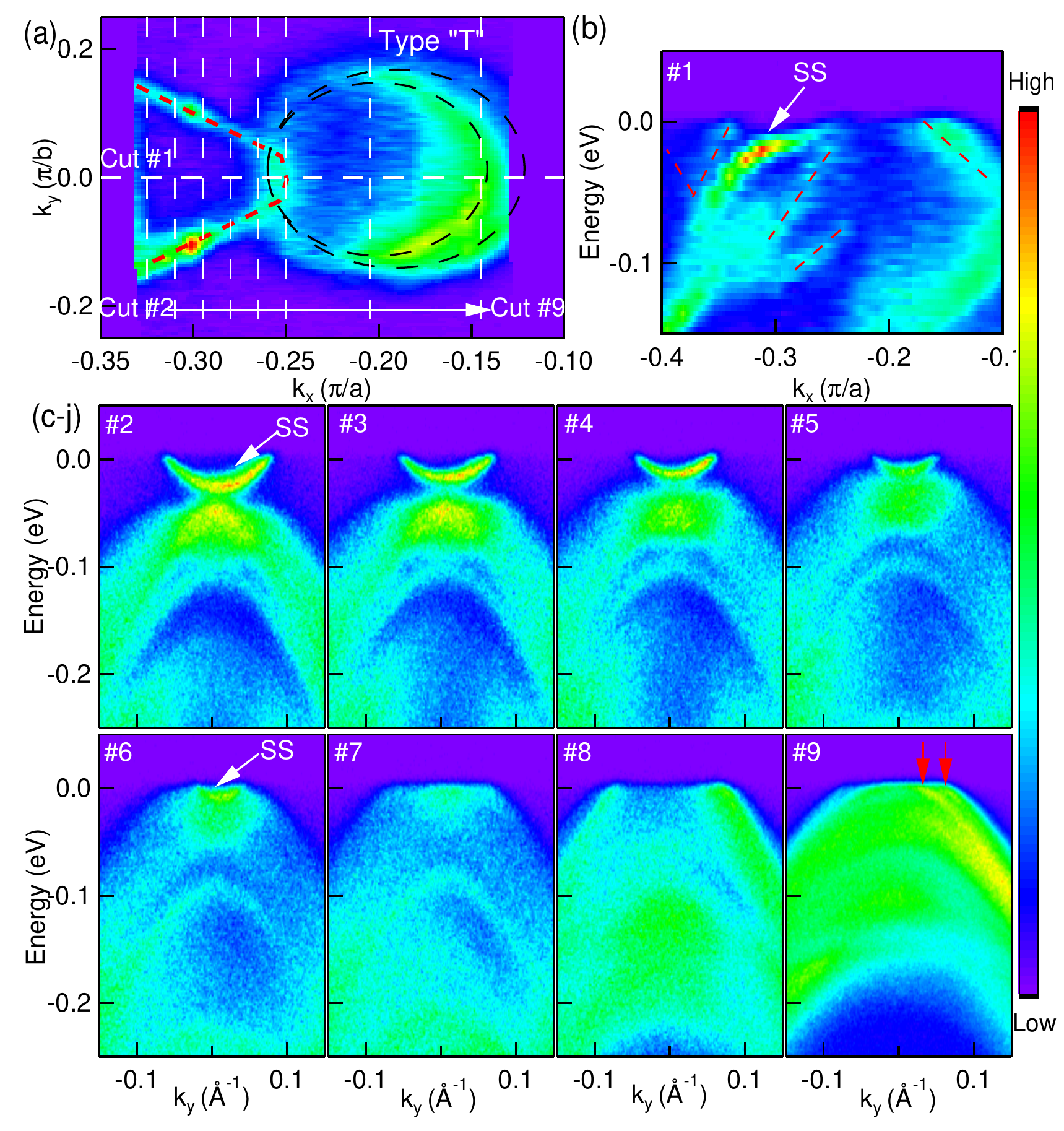}%
	\caption{Fermi surface  and band dispersion measured at $T=16$K and photon energy of 6.7 eV for ``T" type cleave in hole pocket area. 
	(a) Fermi surface plot of ARPES intensity integrated within 5 meV about the chemical potential. Curved red dashed line marks the contour of the Fermi arcs and the black dashed lines mark the contour of the two bulk hole pockets.
	(b)-(j) Band Dispersion along cuts \#1--\#9.  Dashed lines in (b) marked the electron pocket and the two hole pockets. The white arrows point to the location of the band forming the Fermi arc. The red arrows in (j) point to the locations of the two hole bands crossing Fermi level.
	\label{fig:Fig2}}
\end{figure}

\begin{figure}[tb]
	\includegraphics[width=3.2in]{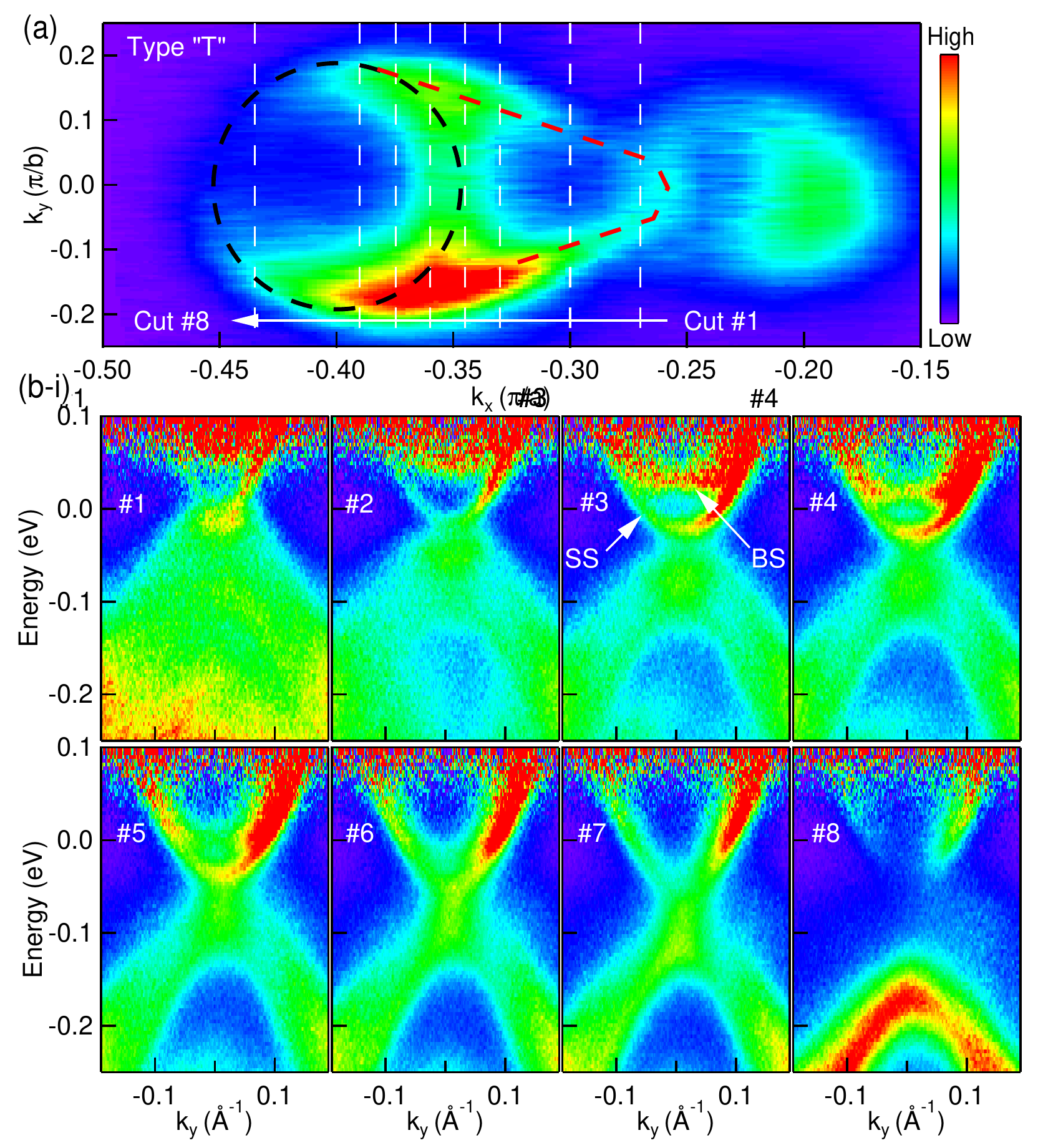}%
	\caption{Fermi surface and band dispersion measured at $T=160$K and photon energy of 6.7 eV for the ``T" type cleave. 
	(a) Fermi surface plot of ARPES intensity integrated within 10 meV about the chemical potential. Curved red dashed line marks the contour of the Fermi arcs and the black dashed line marks the contour of the bulk electron pocket.
	(b)-(i) Band Dispersion along cuts 1-8. The white arrows point to the location of the Fermi arcs and the Bulk State (BS). 
	\label{fig:Fig3}}
\end{figure}

\begin{figure}[tb]
	\includegraphics[width=3.2in]{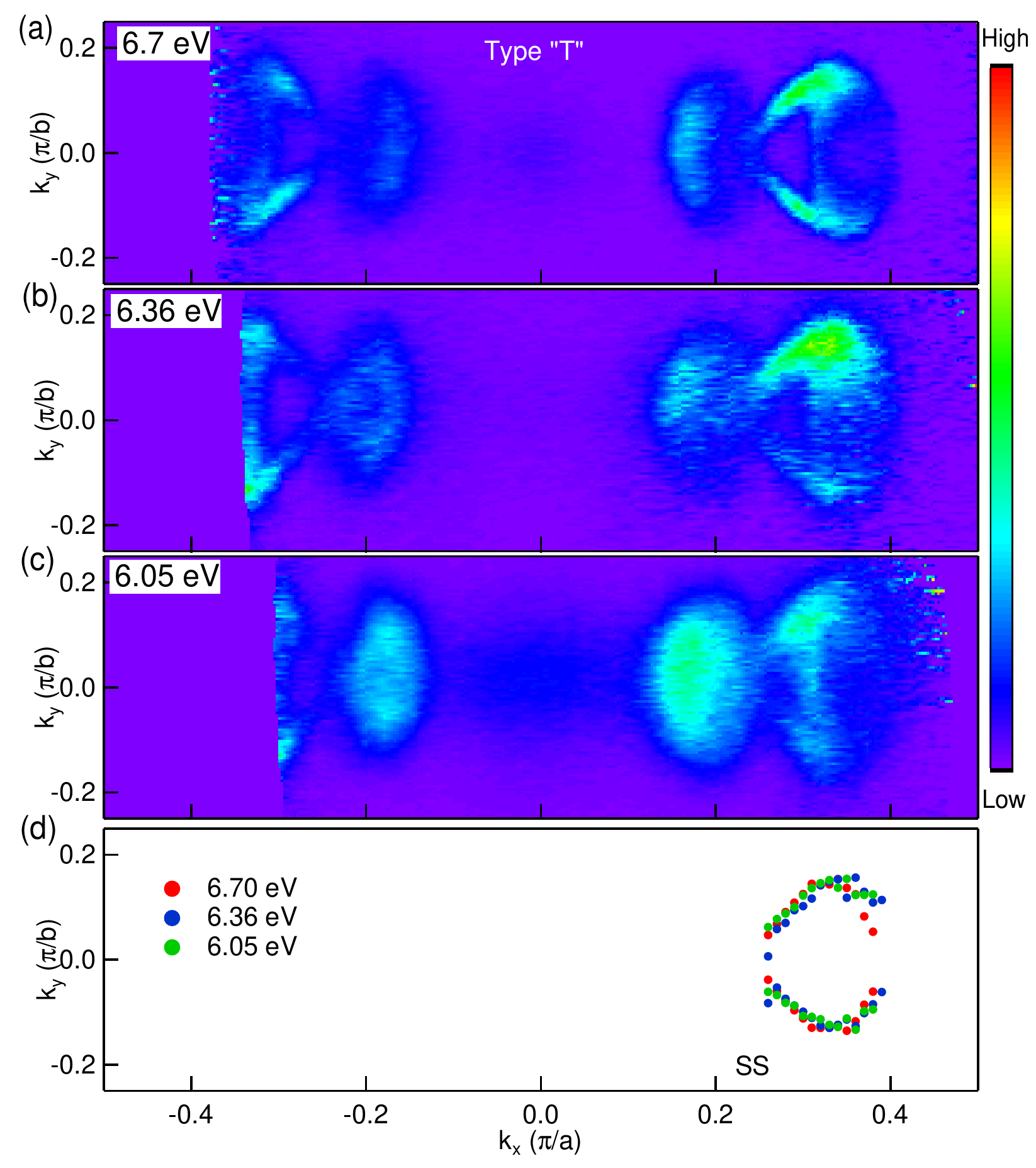}%
	\caption{Fermi surface plot measured at $T=40$K. 
	(a)--(c) Fermi surface plot of ARPES intensity integrated within 10 meV about the chemical potential measured at photon energy of 6.7 eV, 6.36 eV and 6.05 eV, respectively. 
	(d) Fermi surface contour extracted from peak positions of momentum dispersion curves.
	\label{fig:Fig4}}
\end{figure}

Here, we present the study of the electronic structure of WTe$_2$ by using ultrahigh resolution, tunable, VUV laser-based ARPES system. We observed two  distinct types of electronic structure suggesting that there are two possible types of sample cleaves. At this time it is not clear whether there are two different surface terminations, reconstructions or in some cases the cleaving introduces stress at the surface. In type N (normal) cleaves the Fermi surface consists of two pairs of electron pockets and two pairs of hole pockets in agreement with previous studies \cite{Pletikosic14PRL, Wu2015PRL, Jiang2015Signature}. In type T (topological) cleaves we demonstrate that two pairs of Fermi arcs are present that link the electron and hole pockets. This feature is consistent with theory prediction that this material is a host of type II Weyl semimetallic state. 

Whereas most of the previous measurements have been carried out on WTe$_2$ crystals grown via chemical vapor transport using halogens as transport agents \cite{Brown1966, Ali14Nat}, we have grown WTe$_2$ single crystals from a Te-rich binary melt.  High purity, elemental W and Te were placed in alumina crucibles in W$_1$Te$_{99}$ and W$_2$Te$_{98}$ ratios.  The crucibles were sealed in amorphous silica tubes and the ampoules were heated to 1000$^{\circ}$C over 5 hours, held at 1000$^{\circ}$C for 10 hours, and then slowly cooled to 460$^{\circ}$C over 100 hours and finally decanted using a centrifuge \cite{Canfield92PMPB}. The resulting crystals were blade- or ribbon like in morphology with typical dimensions of 3 $\times$ 0.5 $\times$ 0.01 mm with the crystallographic $c$ axis being perpendicular to the crystal surface; the crystals are readily cleaved along this crystal surface. 

Samples were cleaved \textit{in situ} at 16~K and 40~K under ultrahigh vacuum (UHV). The data were acquired using a tunable VUV laser ARPES system, consisting of a Scienta R8000 electron analyzer, picosecond Ti:Sapphire oscillator and fourth harmaonic generator \cite{Jiang14RSI}. Data were collected with a tunable photon energies from 5.3 eV to 6.7 eV. Momentum and energy resolution were set at $\sim$ 0.005 ~\AA$^{-1}$ and 1 meV, respectively. The size of the photon beam on the sample was $\sim$30 $\mu$m.

Two types of Fermi surface and band dispersions along high-symmetry directions in the Brillouin zone (BZ) are shown in Fig.\ref{fig:Fig1}. The Fermi surface of type N cleave is shown in panel (a) and the band dispersion along the $\Gamma$-X symmetry direction is plotted in panel (b) as previously reported \cite{Wu2015PRL}. Fig.\ref{fig:Fig1}(c) shows Fermi surface plot for the type T cleave in the first BZ, integrated within 10 meV about the chemical potential, with the high intensity contours marking the location of the Fermi surface crossings. Presence of Fermi arcs that connect the bulk hole and electron pockets is clearly visible. This is further confirmed by examining the band dispersions along Cuts \#1--\#6 as shown in Figs.\ref{fig:Fig1}(d)--\ref{fig:Fig1}(i). In addition to the two nearly degenerate electron bands and two  branches of the left side of hole pockets (marked by the red dashed lines), a high intensity, sharp band dispersion can be clearly seen, that connects the bottom of the electron pockets and top of the hole pockets. In Figs.\ref{fig:Fig1}(e)--\ref{fig:Fig1}(i), the high symmetry cuts for the type T cleave look almost the same as type N cleaves, except for Cut \#3. Here, an additional electron band is present, demonstrating the existence of an additional electron pocket on the surface. We will examine this feature in more details below.

Fig.\ref{fig:Fig2} shows details of the Fermi sheets and band dispersion of the unusual surface state. In panel (a) we plot the ARPES intensity (integrated within 5 meV about the chemical potential) close to momentum region, where the surface state connects to hole pocket. The curved red dashed line and the black dashed lines mark the outline of the surface electron pockets and two almost degenerate hole pockets, respectively. The band dispersion along Cut \#1 is shown in Fig.\ref{fig:Fig2}(b), where the white arrow points to the location of the surface state. Detailed band dispersions along white vertical cuts are shown in Figs.\ref{fig:Fig2}(c)--\ref{fig:Fig2}(j). The bottom of this surface band dips only slightly below the E$_F$ demonstrating its electron character. This band is much sharper than the lower energy broad, bulk hole bands, consistent with its surface origin. As we move towards the zone center, the electron band shrinks and moves closer to the Fermi level, while the lower hole bands move up. The Fermi arc surface state touches the hole bands at Cut \#5 (panel (f)) and is completely swallowed by the lower hole bands along Cut \#6 (panel (g)). After Cut \#6, the hole bands continue moving up and finally cross the Fermi level and forms a pair of hole pockets. We can clearly see the separation of the almost degenerate hole pockets along Cut \#9 (panel (j)), as marked by the red arrows pointing at the crossing points. 

The merging between Fermi arcs and bulk electron pockets is shown in Fig.\ref{fig:Fig3}. Panel (a) shows the ARPES intensity integrated within 10 meV about the chemical potential and measured at $T=160$K. The black dashed line and red dashed line mark the location of the bulk electron pockets and Fermi arc band, respectively. In order to better show the details of the bulk electron pockets, we have plotted the ARPES intensity divided by the Fermi function along the white vertical cuts \#1--\#8 in Figs.\ref{fig:Fig3}(b)--\ref{fig:Fig3}(i). At Cut \#1 (panel (b)), a single electron pocket is clearly observed that touches the top of the lower hole bands and forms the beginning of two Fermi  arcs on either side.  As we move away from the hole pockets, the band responsible for Fermi arcs moves to higher binding energy. Slightly further (cut \#2) a bulk band becomes visible still above the E$_F$. Both bands are very clearly visible starting from cut \#3, where they are indicated by white arrows and labelled.  Closer to the center of the bulk electron pocket, the two bands eventually merge together. The detailed band evolution in Fig.\ref{fig:Fig2} and Fig.\ref{fig:Fig3} demonstrates that the Fermi arc states connect the bottom of the electron pockets and top of the hole pockets, consistent with the previous theoretical prediction \cite{Soluyanov2015Type}.

To verify the surface origin of Fermi arc band, we have carried out photon energy dependent measurements and present them in Fig.\ref{fig:Fig4}. Panels (a)--(c) show the ARPES intensity integrated within 10 meV about the chemical potential measured at photon energies of 6.7 eV, 6.36 eV and 6.05 eV, respectively. We can clearly see that the shape of the bulk electron pockets and hole pockets change slightly with photon energy, but the contour of the Fermi arcs does not. To better quantify our results, we have plotted the contour obtain by fitting to the data of the Fermi arc band in Fig.\ref{fig:Fig4}(d). The momentum locations of the Fermi arcs and its shape remains the same for all three photon energies. This is consistent with behavior expected of a Weyl semimetal, where the Fermi arcs connect the projection of the 3D Weyl points.

In summary, we have used ultrahigh resolution, tunable, laser-based ARPES to study the electronic properties of WTe$_2$, a compound that was predicted to be a type-II Weyl semimetal. We found two different cleave types that have distinct electronic structure. First type is consistent with previous studies, while the second type displays clear Fermi arcs that connect the hole and electron pockets. As discussed previously \cite{Soluyanov2015Type, Chang15Arc, Huang2016Spectroscopic}, the electronic structure of WTe$_2$ and MoTe$_2$ may change even significantly if a small change in the lattice parameters (e. g. strain) is considered in band structure calculations. This is a possible explanation of the two types of observed electronic structures. The Fermi arcs reported here are long sought after signatures of the type-II Weyl semimetallic state that were predicted theoretically.

We would like to thank Nandini Trivedi and Tim McCormick for very useful discussions. Research was supported by the U.S. Department of Energy, Office of Basic Energy Sciences, Division of Materials Sciences and Engineering. Ames Laboratory is operated for the U.S. Department of Energy by the Iowa State University under Contract No. DE-AC02-07CH11358. N.H.J. is supported by the Gordon and Betty Moore Foundation EPiQS Initiative (Grant No. GBMF4411). L.H. was supported by CEM, a NSF MRSEC, under Grant No. DMR-1420451.

Note added: After completion of this work we become aware of similar results \cite{Bruno2016Surface, Wang2016Spectroscopic} that are in good agreement with our conclusions.

\bibliography{WTe2}

\end{document}